\documentclass[aps,prl,preprint,superscriptaddress]{revtex4-1}
\usepackage{graphicx}  
\usepackage{dcolumn}   
\usepackage{bm}        
\usepackage{amssymb}
\usepackage[english]{babel}
\usepackage{booktabs}
\usepackage[export]{adjustbox}
\usepackage{array}
\usepackage{subcaption}
\usepackage{hyperref}
\hypersetup{
	colorlinks=true,
	linkcolor=blue,
	filecolor=magenta,      
	urlcolor=blue,
}
\begin{document}
\title{Pressure induced emergence of visible luminescence in $Cs_3Bi_2Br_9$: Effect of structural distortion in optical behaviour}

\author{Debabrata Samanta}
\affiliation{National Centre for High Pressure Studies, Department of Physical Sciences, Indian Institute of Science Education and Research Kolkata, Mohanpur Campus, Mohanpur 741246, Nadia, West Bengal, India.}

\author{Pinku Saha}
\affiliation{National Centre for High Pressure Studies, Department of Physical Sciences, Indian Institute of Science Education and Research Kolkata, Mohanpur Campus, Mohanpur 741246, Nadia, West Bengal, India.}

\author{Bishnupada Ghosh}
\affiliation{National Centre for High Pressure Studies, Department of Physical Sciences, Indian Institute of Science Education and Research Kolkata, Mohanpur Campus, Mohanpur 741246, Nadia, West Bengal, India.}
\author{Sonu Pratap Chaudhary}
\affiliation{Department of Chemical Sciences, Indian Institute of Science Education and Research Kolkata, Mohanpur Campus, Mohanpur 741246, Nadia, West Bengal, India.}

\author{Sayan Bhattacharyya}
\affiliation{Department of Chemical Sciences, Indian Institute of Science Education and Research Kolkata, Mohanpur Campus, Mohanpur 741246, Nadia, West Bengal, India.}

\author{Swastika Chatterjee}
\affiliation{Department of Earth Sciences, Indian Institute of Science Education and Research Kolkata, Mohanpur Campus, Mohanpur 741246, Nadia, West Bengal, India.}

\author{Goutam Dev Mukherjee}
\email [Corresponding author:]{goutamdev@iiserkol.ac.in}
\affiliation{National Centre for High Pressure Studies, Department of Physical Sciences, Indian Institute of Science Education and Research Kolkata, Mohanpur Campus, Mohanpur 741246, Nadia, West Bengal, India.}
\date{\today}
\begin{abstract} 
We report emergence of photoluminescence at room temperature in trigonal $Cs_3Bi_2Br_9$ at high pressures. Enhancement in intensity with pressure is found to be driven by increase in distortion of $BiBr_6$ octahedra and iso-structural transitions. Electronic band structure calculations show the sample in the high pressure phase to be an indirect band gap semiconductor. The luminescence peak profile show signatures related to the recombination of free and self trapped excitons, respectively. Blue shift of the both peaks till about 4.4 GPa are due to the exciton recombination before relaxation due to the decrease in exciton lifetime with scattering from phonons
\end{abstract}
\maketitle


In recent years, halide perovskites of the formula $A_3B_2X_9$ ($A$:Cs, Rb; $B$:Sb or Bi; and $X$:halogen) have been in research forefront due to their potential applications in optoelectronic devices and solar cells\cite{Anna,Fan,Bin}. Since toxicity of lead has been a consistent concern for the environment and health safety, lead free halide perovskites have a great importance. Among several important materials under study, $Cs_3Bi_2Br_9$ has emerged as a model system for  detailed investigation as it has been found to be stable under atmospheric conditions. At ambient conditions $Cs_3Bi_2Br_9$  has trigonal crystal structure with space group $P\bar{3}m1$ and has been found to be ferroelectric\cite{Lazarini}. $Cs_3Bi_2Br_9$ is found to undergo a structural transition to monoclinic phase at 95 K \cite{Bator,IP}. Early studies on $Cs_3Bi_2Br_9$ single crystals by Timmermans and Blasse \cite{cwm} showed blue luminescence at 4K, which got quenched by 20K. A recent careful measurements by Bass et al.\cite{bass} on high purity nanocrystalline $Cs_3Bi_2Br_9$ show a weak but structured photoluminescence (PL) at room temperature and their feature is explained by excitons coupling to vibrons in the lattice.  All the above studies indicate large exciton binding energy leading to radiative recombination of excitons.

Pressure, being a unique thermodynamic variable, may dramatically alter the crystal structure, and hence optical properties of materials. High pressure (HP) studies on $CsPbBr_3$ nanocrystals have led to new structures with enhanced luminescence properties \cite{Yasutaka}. Ma et al.\cite{Zhiwei} have shown that by pressure tuning, initially non-luminescence $Cs_4PbBr_6$ nanocrystals exhibit a blue luminescence above 3 GPa  due to a structural transition. 
The unit cell of $Cs_3Bi_2Br_9$ has a lot of empty space as one third of octahedral positions are vacant. Therefore, pressure is expected have a significant effect on  its structural behaviour. Recently Li et al.\cite{Qian} have reported pressure induced structural transition of 2D hybrid halide perovskite $(CH_3NH_3)_3Bi_2Br_9 $. In this work we have carried out a detailed pressure dependent study on structural, and optical  properties of $Cs_3Bi_2Br_9$. We observe an anomalous increase in PL intensity at about 2.9 GPa mediated by distortion and elongation of $BiBr_6$ octahedra due to iso-structural transition induced by high compressibility of the sample.

The sample is synthesized in the laboratory using sol-gel technique (details are given in Supplementary Information (SI)). Characterization of synthesized sample using x-ray diffraction (XRD) revealed trigonal structure ($P\bar{3}m1$) with lattice parameters: a=7.9689(2)$\AA$, c=9.8577(6)$\AA$, and are in good agreement with previous report \cite{Lazarini}. 
We have recorded HP PL spectra of $Cs_3Bi_2Br_9$ using excitation wavelength of 488nm and shown in Fig.1. $Cs_3Bi_2Br_9$ does not show any observable PL at ambient condition. 
Interestingly with increase in pressure, a luminescence appears at about 1.4 GPa as shown in the bottom of Fig.1(a). The intensity is too low compared to the ambient background. The PL intensity shows a maximum at about 2.9 GPa and then almost disappears above 11 GPa. The PL peak profiles show a long tail at lower energies with a kink at higher energy side(SI Fig.S2).  
In Fig.1(b), we have shown the pressure evolution of peak positions of the two peaks. Both the peaks show a blue-shift till about 4.4 GPa followed by a redshift. Fig.1(c) shows the relative change in integrated intensities of both the peaks. The higher energy Peak-A shows much larger change, though its intensity is much lower compared to Peak-B. We shall discuss the results in later sections in detail after we discuss the structural evolution with pressure.

In Fig.2(a) we have shown the pressure evolution of Raman spectra (excitation $\lambda$ = 532 nm) of $Cs_3Bi_2Br_9$ up to about 16 GPa. 
At ambient condition we observe six Raman modes (SI Fig.S3). The modes at $167cm^{-1}$ and $192cm^{-1}$ are indexed to $E_g$ and  $A_{1g}$, respectively and are associated with stretching vibrations of $Bi-Br$ bonds \cite{Bator}. Other four are labeled as $M_1$($63.9cm^{-1}$), $M_2$($67.3cm^{-1}$), $M_3$($78.8cm^{-1}$), $M_4$($92.2cm^{-1}$) and originate from deformation of $BiBr_6$ octahedra due to vibrations of Br atoms only\cite{ Bator, MYa}. 
With pressure all modes are blue shifted. The first four low frequency modes broaden with pressure with reordering of their respective intensities.
At 4.7 GPa $E_g$, $A_{1g}$ modes split into two modes each (labeled as $M_5$, $M_6$, $M_7$, $M_8$) indicating a structural transition with lowering in symmetry. Gradual extensive broadening of all Raman modes with pressure indicate to amorphization of the sample.
Pressure evolution of full width half maximum (FWHM) of $A_{1g}$, $E_g$ are shown in Fig.2(b). Increase in the life time of both the phonon modes is evident from sharp decrease of FWHM till 2.4 GPa. Similarly in Fig.2(c) one can see an increase in the intensity of symmetric stretching mode $A_{1g}$ with respect to that of asymmetric stretching mode $E_g$.  Enhancement in a Raman mode intensity represents a strong coupling between the excited electrons and particular vibrational mode of the lattice as well as enhancement in the dipole-dipole polarizability of the lattice. The above anomalous behaviour of the Raman modes and enhancement of PL do indicate a strong exciton-phonon coupling.  

We have carried out HP XRD measurements up to about 25 GPa on powder $Cs_3Bi_2Br_9$ at room temperature.  
Pressure evolution of XRD patterns show appearance and disappearance of certain Bragg peaks (Fig.3(a)). We assign the ambient phase as Trigonal(I). At about 1.3 GPa a new peak appears at $2\theta$ = 8.72$^{\circ}$. Indexing results in a trigonal structure with lattice parameters: $a$=7.7350(2), $c$=28.934(1) $\AA$; space group $P\bar{3}m1$ (Trigonal(II)). In Trigonal(II) phase the $c$-axis is almost three times of that  of  Trigonal(I) phase.  With further increase in pressure to 3.1 GPa, another new Bragg peak appears at $2\theta$ = 9.93$^{\circ}$. The XRD pattern is indexed again to trigonal structure with lattice parameters: $a$=15.0956(4), $c$=9.4438(4)$\AA$; space group $P\bar{3}m1$ (Trigonal(III)) with $a$ and $b$-axes being double with respect to that of Trigonal(I) phase. Several new peaks emerge at 6 GPa indicating lowering in symmetry and the XRD pattern is indexed to a triclinic phase with structural parameters: $a$=9.455(1), $b$=7.4426(9), c=14.168(2) $\AA$, $\alpha$=59.160(7)$^{\circ}$, $\beta$=105.635(7)$^{\circ}$, $\gamma$=98.549(7)$^{\circ}$; space group $P\bar{1}$. Therefore the XRD data show two iso-structural transitions followed by a change in symmetry. The broadening of Bragg peaks and their disappearance above 11 GPa corroborate the pressure induced amorphization of the sample. Below 10 GPa, the sharp Bragg peaks of the Ag-pressure marker show the absence of non-hydrostatic stress on the sample. $3^{rd}$ order Birch-Murnaghan EOS is fitted to the volume {\it vs.} pressure data by normalizing the volume of all the trigonal phases with respect to the ambient phase (SI Fig.S6). The EOS fit results in low bulk moduli ($B_0$) values: $12\pm 1$ and $61.5\pm0.6$ GPa for trigonal and  triclinic phases, respectively. 
Rietveld refinement of all trigonal phases are carried out starting with the model of the Trigonal(I) phase \cite{Lazarini}. The refined atom positions and XRD patterns are documented in the SI (Table-SI, SII, SIII and Fig.S7). 
In Fig.3(b) we have shown the octahedral arrangement in the unit cell for all trigonal phases. Trigonal(II) and Trigonal(III) phase consists of  three and two types $BiBr_6$ octahedra, respectively. The large compressibility results in a increase of distortion index (DI) by about 67\% and quadratic elongation (QE) by about 1\% of $BiBr_6$ octahedra in the pressure range 0-0.9 GPa itself( SI Table-SIV).
It can be seen that Bi-atoms have moved significantly away from the central positions of the octahedra resulting in different $Bi-Br$ bond lengths and bond angles giving rise to significant rotation and distortion of the octahedra. The increase in distortion in $BiBr_6$ octahedra leads to the anomalous behaviour of Raman modes.
 
Using first principles DFT calculations we have computed the PBE+SOC band structure of $Cs_3Bi_2Br_9$ at  0.1 and 3.1GPa. Electronic band structure and partial density of states (PDOS) predicted by our calculations for the 0.1 GPa structure agree very well with those calculated using hybrid functionals (HSE06+SOC) for the same system at ambient  pressure\cite{bass} (SI Fig.S8). 
It is to be noted that PBE is known to under-estimate the band-gap and a more accurate calculation involving computationally far more expensive hybrid functional is expected to give a better estimate of the gap.  Nonetheless, our PBE+SOC calculations are expected to provide a good qualitative picture of the nature of the band-structure of the material close to the Fermi level.
Fig.4 shows the PBE+SOC band structure and PDOS of this material at 3.1 GPa, calculated using our experimentally determined crystal structure. The band structure shows that the system is an indirect band gap semiconductor, with none of the low lying excitations occur at the same $k$-point. We find that the top of the valence band is dominated by $Br-p$ and $Bi-s$, whereas the bottom of the conduction band is dominated by $Bi-p$ indicating the importance of Bi-centers in the $BiBr_6$ octahedra for photoinduced activity. Also the electronic  band gap(3.1 GPa) is found to get reduced with increase in pressure.

The anomalous changes observed in Raman spectroscopy and XRD studies do indicate that application of pressure leads to several structural changes induced by the deformation of $BiBr_6$ polyhedra  affecting the electronic properties. The emergence of PL can be related to the phase transition  with its appearance at Trigonal(I) - Trigonal(II) transition. The same can be corroborated by  the fact that  the PL intensity reaches its maximum at about 2.9 GPa (Trigonal-III phase) and shows a $2^{nd}$ maximum at about 4.4 GPa(Triclinic phase) (Fig.1(b)). The PL peak feature shows a structure with two peaks suggesting coexistence of free exciton(FE) and self-trapped exciton (STE) recombination\cite{Shunran, Jiangjian, Ivan}. We have carried out an analysis of the obtained PL intensity ($I$) by changing the excitation laser power ($L$) using the power law behaviour $I\sim L^K$ law\cite{Schmidt} ({\bf SI Fig.S9}).The value of $K$ from the fit is found to be 1.17 for Peak-A and 1.04 for Peak-B. Schimdt and Lischka \cite{Schmidt} showed that for free and bound excitons the value of $K$ should be: $1<K<2$. This confirms the excitonic nature of pressure induced luminescence in $Cs_3Bi_2Br_9$. Peak-A with its narrow feature can be the free exciton. Iso-structural phase transition driven by structural distortion of $BiBr_6$ octahedra can give rise to photoinduced pseudo Jahn-Teller distortion of the individual $BiBr_6$ octahedra and  lead to the formation and radiative recombination of STE. The same is reflected in the asymmetric nature of Peak-B profile with a long tail in low-energy side. The initial blue shift of the PL peaks till about 4.4 GPa also can be related to the extensive structural distortion before transition to lowest symmetry triclinic phase. Due to the indirect band gap of the sample, the life time of the excitons shortens because of scattering with one or more phonon and recombination of exciton happens much before exciton relaxation. The red-shift of the PL peaks above 4.2 GPa is probably due to decrease of band gap \cite{Takuyao}. This shows that the structural distortion is a important factor to be considered for lead-free halide perovskites for exhibiting luminescence. We believe that this work will lead to further studies on Pb-free stable inorganic halide perovskites.

\section{Conclusions}
We have carried out a detailed high pressure study on $Cs_3Bi_2Br_9$ perovskite by PL, XRD and Raman spectroscopy measurements. HP XRD and Raman measurements show two isostructural transitions in ambient trigonal phase at about 0.9 GPa and 2.4 GPa, respectively. These isostructural transitions take place with extensive structural distortion due to large compressibility of the system. The sample shows appearance of PL at about 1.4 GPa driven by the isostructural transition. PL feature shows a kink at about 2.1 eV due to free exciton recombination and another broad peak at about 1.8 eV due to self trapped exciton (STE) recombination. Electronic density of states and band structure calculation show  that the valence band and conduction band consist of hybridization of $Br-p$, $Bi-s$ orbitals and $Bi-p$ orbitals. The enhancement of PL intensity and blue shift in PL peaks are related to the large distortion and quadratic elongation of $BiBr_6$ octahedra.

\section{Acknowledgments}
We acknowledge the financial support from the Ministry of Earth Sciences, Government of India, grant number MoES/16/25/10-RDEAS. The financial support from Department of Science and Technology, Government of India under Indo-Italy Executive Programme of Scientific and Technological Cooperation is gratefully acknowledged. DS, PS and BG also acknowledge the fellowship grant supported by the INSPIRE program, Department of Science and Technology, Government of India.


\begin{figure}
\centering     
\includegraphics[width=0.6\columnwidth]{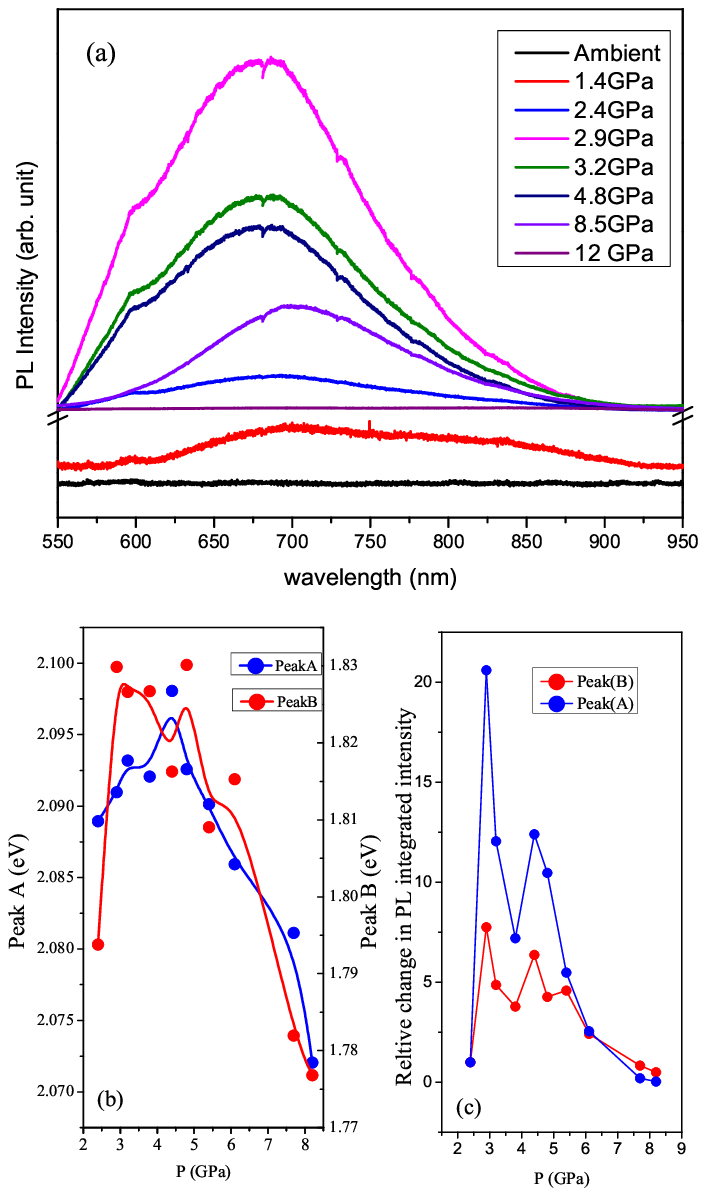}




\caption{Pressure evolution of (a) photoluminescence pattern of $Cs_3Bi_2Br_9$; (b) PL peak positions; (c) relative change in integrated intensity of both PL peaks.}
\end{figure}

\begin{figure}
\centering     
\includegraphics[width=0.6\columnwidth]{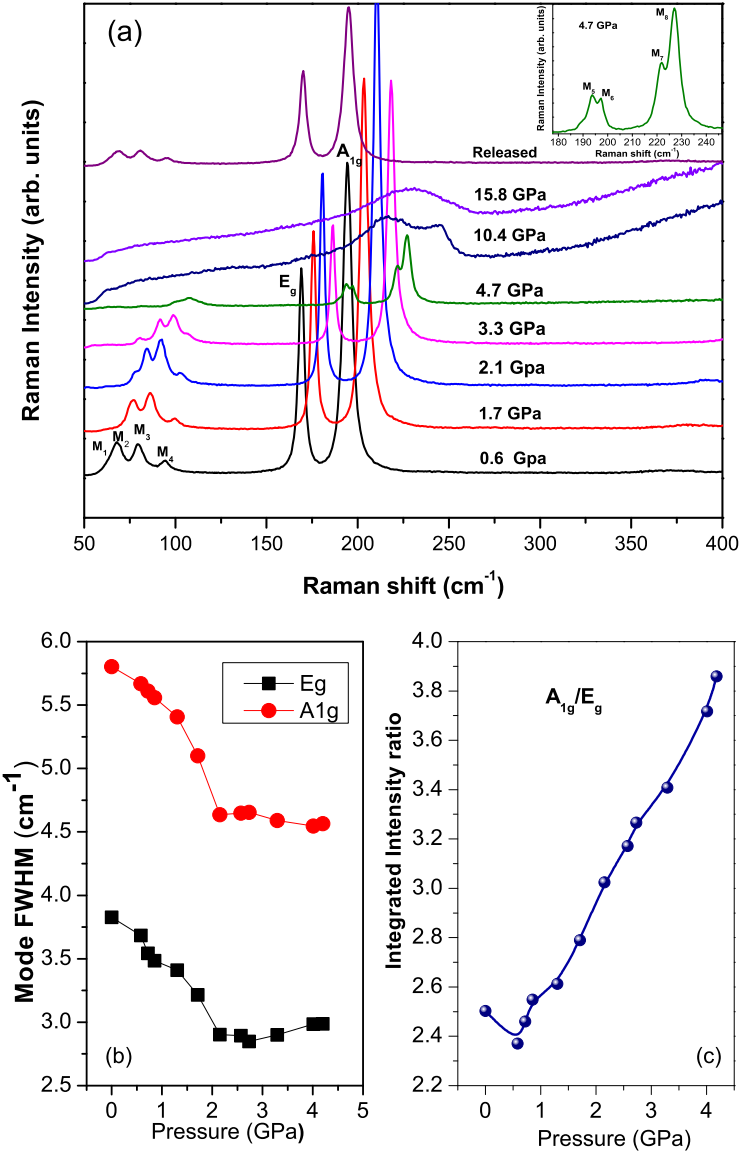}




\caption{(a) Pressure evolution of Raman spectra of $Cs_3Bi_2Br_9$. The inset shows the splitting of $A_{1g}$ and $E_g$ modes at 4.7 GPa. Pressure evolution of (b) FWHM of $A_{1g}$ and $E_g$ modes and (c) relative change in intensity of $A_{1g}$ with respect to $E_g$.}
\end{figure}

\begin{figure}
\centering     
{\includegraphics[width=0.8\columnwidth]{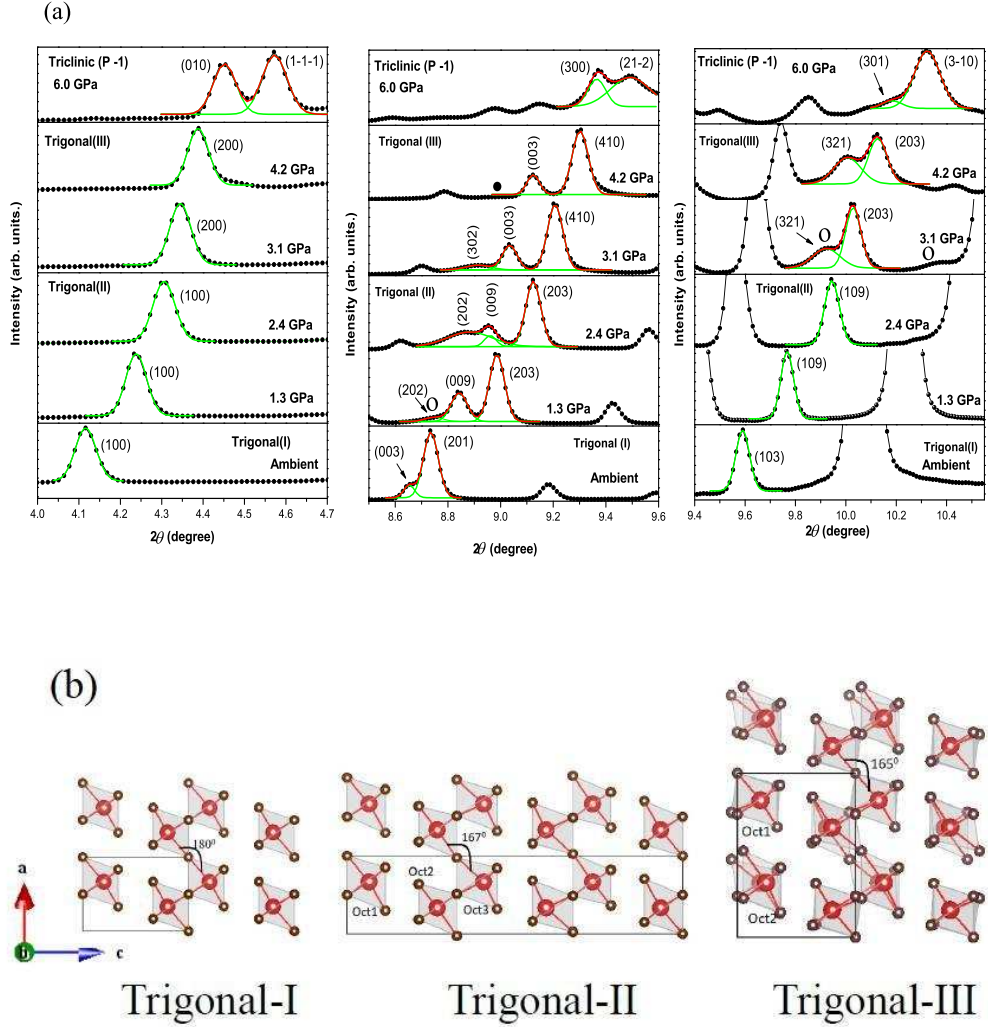}}\\
\caption{(a) Evolution of XRD patterns at selected pressure points. Splitting, disappearance (dark circles) and emergence (open circles) of Bragg peaks  show the structural transitions with pressure. (b) Arrangements of $BiBr_6$ octahedra in three trigonal phases. The $BiBr_6$ octahedra show large distortions in Trigonal(II) phase with the angle $Bi-Br-Bi$ connecting different octahedra at the corrugated junctions decreasing with pressure. In Trigonal(III) phase the rotation of octahedra along b-axis with respect to each other depicts large disorder in the unit cell.}
\end{figure}


\begin{figure}
\centering     
{\includegraphics[width=\columnwidth]{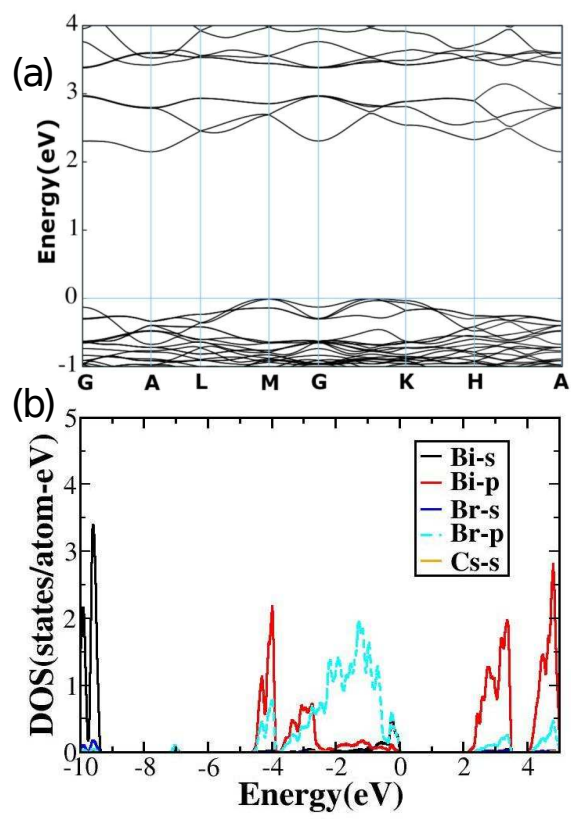}}
\caption{(a) Electronic band structure of and (b) electronic partial density of states of $Cs_3Bi_2Br_9$ at 3.1 GPa (Trigonal(III) phase).  }
\end{figure}
\end{document}


\title{Supplementary Information for Pressure induced emergence of visible luminescence in $Cs_3Bi_2Br_9$: Effect of structural distortion in optical behaviour}

\author{Debabrata Samanta}
\affiliation{Department of Physical Sciences, Indian Institute of Science Education and Research Kolkata, Mohanpur Campus, Mohanpur 741246, Nadia, West Bengal, India.}

\author{Pinku Saha}
\affiliation{Department of Physical Sciences, Indian Institute of Science Education and Research Kolkata, Mohanpur Campus, Mohanpur 741246, Nadia, West Bengal, India.}

\author{Bishnupada Ghosh}
\affiliation{Department of Physical Sciences, Indian Institute of Science Education and Research Kolkata, Mohanpur Campus, Mohanpur 741246, Nadia, West Bengal, India.}
\author{Sonu Pratap Chaudhary}
\affiliation{Department of Chemical Sciences, Indian Institute of Science Education and Research Kolkata, Mohanpur Campus, Mohanpur 741246, Nadia, West Bengal, India.}

\author{Sayan Bhattacharyya}
\affiliation{Department of Chemical Sciences, Indian Institute of Science Education and Research Kolkata, Mohanpur Campus, Mohanpur 741246, Nadia, West Bengal, India.}

\author{Swastika Chatterjee}
\affiliation{Department of Earth Sciences, Indian Institute of Science Education and Research Kolkata, Mohanpur Campus, Mohanpur 741246, Nadia, West Bengal, India.}

\author{Goutam Dev Mukherjee}
\email [Corresponding author:]{goutamdev@iiserkol.ac.in}
\affiliation{Department of Physical Sciences, Indian Institute of Science Education and Research Kolkata, Mohanpur Campus, Mohanpur 741246, Nadia, West Bengal, India.}
\date{\today}

\begin{abstract}
This document contains details of sample preparation, experimental methods and theoretical simulation methods. Also additional tables and figures are given documenting our analysis details.
	\end{abstract}
	
	\maketitle

\section{Sample preparation}

Crystalline powders of $Cs_3Bi_2Br_9$ are prepared by dissolving 63.84mg of $CsBr$ and 89.74mg of $BiBr_3(\geq99\%)$ with $3:2$ molar ratio in 2 mL of $HBr(99.999\%)$ (48 wt. $\%$ in water $\geq$ 99.99 $\%$) in a capped vial and then the mixture are heated to $110^{o}$C for 2 hours. After that, the mixture are kept for an additional 2 hours at the same temperature without stirring. The obtained solution is allowed to cool to room temperature overnight, followed by washing with ethanol and drying.
 
\section{Experimental methods}
High pressure (HP) experiments at room temperature are performed using piston cylinder type diamond anvil cell (DAC) of culet size 300$ \mu$m. A small amount of powder sample along with pressure marker are loaded inside the central hole of diameter 100$\mu$m of steel gasket preindented to a thickness 60$\mu$m. For high pressure XRD and Raman measurements 4:1 methanol ethanol mixture are used as a pressure transmitting medium (PTM), whereas for PL measurements silicon oil is used. PL and Raman spectroscopy measurements are carried out using a confocal micro-Raman spectrometer(Monovista from SI GmBH) and Horiba Jobin-Yvon LabRAM HR-800 spectrometer with back scattering geometry respectively. 
A micron sized ruby is loaded along with the sample in the DAC for determination of pressure using ruby fluorescence technique\cite{Mao}.
488nm and 532nm lasers are used to excite the sample for PL and Raman measurements, respectively. 
Infinity corrected 20X objective having long working distance is used to focus the laser and collect the radiation.
Pressure dependent XRD measurements are carried out at  XPRESS beamline in the Elettra synchrotron radiation source using a monochromatic x-ray of wavelength 0.4957$\AA$ collimated to 20 $mu$m at the sample.  We have used $Ag$ as pressure marker and pressure is calculated using  equation of state (EOS) of $Ag$ \cite{Dewaele}. 
FIT2D software\cite{Hammersley} is employed to convert 2D diffraction image to intensity versus 2$\theta$ plot.
All the XRD data are analyzed by CRYSFIRE\cite{Shirley} and GSAS \cite{Brian} program. 
We have used VESTA\cite{Koichi} software for visualization of crystal structure. UV-absorption spectra are recorded using Jasco V-670 spectrophotometer.

\section{THEORETICAL METHODS}

First principles calculations are performed within the framework of density functional theory(DFT)\cite{Hohenberg, Sham} using projected augmented wave(PAW) \cite{PE, Kresse} method as implemented in the plane-wave based VASP code \cite{Hafner, Furth, JFurth}. The exchange-correlation functional is chosen to be the Perdew-Burke-Ernzerhof(PBE)\cite{John} implementation of the generalized gradient approximation (GGA). Spin-orbit coupling(SOC), which is crucial for understanding the electronic structure of the Bi-containing material has been included self-consistently in our calculation. An energy cut-off of 450 eV is used for plane wave expansions. A k-mesh of $9\times9\times9$ is used for the 0.1 GPa structure that contains 14 atoms. Whereas, a k-mesh of $3\times3\times5$ is used for the 3.1GPa crystal structure containing 56 atoms. The lattice parameters have been fixed at experimental values and the ionic positions have been relaxed using conjugate gradient algorithm, until the Helmann-Feynmann forces became less than 0.005 eV/$\AA$. The energy convergence with respect to all computational parameters have been carefully examined.

The upper panel of the Supplementary Fig.S8 shows the computed PBE+SOC band structure at 0.1GPa.  $Cs_3Bi_2Br_9$ has both direct(2.52eV) and indirect (2.4eV) band gap. Direct energy gap is at G whereas indirect energy gap takes place from G to A. It is to be noted that PBE is known to under-estimate the band-gap and a more accurate calculation involving computationally far more expensive hybrid functional is expected to give a better estimate of the gap. Nonetheless, our PBE+SOC calculations are expected to provide a good qualitative picture of the nature of the band-structure of the material close to the Fermi level. This is ascertained by the fact that the nature of the transitions as predicted by our calculations for the 0.1 GPa structure agree very well with those calculated using hybrid functionals (HSE06+SOC) for the same system at ambient  pressure\cite{bass}.  The calculated partial density of states (lower panel of Supplementary Fig.S8) shows that the top of the valence band is dominated by $Br-p$ and $Bi-s$ hybridized orbitals, whereas the bottom of the conduction band is dominated by $Bi-p$.

\begin{table}[h]
	\caption*{Table-SI}{Wyckoff positions, fractional coordinates for the atoms in $Cs_3Bi_2Br_9$ at ambient condition.} 
	
	\begin{tabular}{c c c c c }
		\hline\hline 
		Atom & Wyckoff position & x & y  & z \\ 
		
		\hline 
		Bi1& 2d &0.666667  &   0.333333   &   0.183096   \\
		Cs1&  1a & 0 & 0 &0 \\
		Cs2& 2d & 0.666667  &    0.333333    &  0.671253  \\
		Br1& 3e & 0.5  & 0.5& 0\\
		Br2& 6i & 0.338988   &   0.169493  &    0.333600 \\ 
		\hline\hline
	\end{tabular}
\end{table}

\begin{table}[ht]
	\caption*{Table-SII}{Wyckoff positions, fractional coordinates for the atoms in $Cs_3Bi_2Br_9$ at 2.4GPa.} 
	
	\begin{tabular}{c c c c c }
		\hline\hline 
		Atom & Wyckoff position & x & y  & z \\ [0.3ex]
		
		\hline 
		Bi1& 2d &  0.333330   &   0.666700    &  0.584502  \\
		Bi2&  2d & 0.333300   &   0.666700   &   0.936100  \\
		Bi3& 2d & 0.333300  &    0.666700    &  0.270496  \\
		Cs1& 1a & 0 & 0& 0\\
		Cs2& 2c & 0   &   0   &   0.355505 \\
		Cs3& 2d & 0.333300    &  0.666700   &   0.445541\\
		Cs4& 2d & 0.333300   &   0.666700   &   0.770552 \\
		Cs5& 2d &0.333300  &    0.666700    &  0.111000\\
		Br1& 3e & 0.5   &   0  &    0 \\
		Br2& 6i &  0.541210    & -0.541210    &  0.319422 \\
		Br3& 6i & 0.334367  &    0.167184  &    0.124581  \\
		Br4& 6i & 0.314264    &  0.157132   &   0.777384 \\
		Br5& 6i & 0.295864    &  0.147932   &   0.451950  \\ 
		\hline\hline
	\end{tabular}
\end{table}

\begin{table}[h]
	\caption*{Table-SIII}{Wyckoff positions, fractional coordinates for the atoms in $Cs_3Bi_2Br_9$ at 3.1GPa.} 
	
	\begin{tabular}{c c c c c }
		\hline\hline 
		Atom & Wyckoff position & x & y  & z \\ 
		
		\hline 
		Bi1& 2d &  0.333300   &   0.666700    &  0.243004  \\
		Bi2&  6i & 0.834698     & 0.165302    &  0.196654  \\
		Cs1& 1a & 0 & 0& 0\\
		Cs2& 3e & 0.5   &   0   &   0 \\
		Cs3& 2d & 0.333300    &  0.666700    &  0.674238 \\
		Cs4& 6i & 0.837460   &   0.162540  &    0.663645  \\
		Br1& 6g & 0.25   &   0  &    0 \\
		Br2& 6i &   0.762906   &   0.237094    & -0.050717   \\
		Br3& 6i & 0.927347   &   0.072653   &   0.361501   \\
		Br4& 6i & 0.424332    &  0.575668   &   0.366233  \\
		Br5& 12j & 0.671844    &  0.079077   &   0.335866   \\  
		\hline\hline
	\end{tabular}
\end{table}

\begin{table}[ht]
	\caption*{Table-SIV}{Distortion index and quadratic elongation of octahedra at in different phases. Multiple rows in Trigonal(II) and Trigonal(III) phases are due to the fact that Trigonal(II) phase contains three different type of octahedra and Trigonal(III) phase ontains two different type of octahedra.} 
	\centering 
	\begin{tabular}{c c c c}
		\hline\hline 
		Phase & Pressure\,\,\,\, & Distortion index \,\,\,\,& Quadratic elongation \\
		      & GPa      &                &    \\
		\hline 
		Trigonal(I)& 0  & 0.03923&1.0035 \\
							& 0.9 & 0.06503&1.0133\\
		\hline
		\multirow{6}{*}{Trigonal(II)}&\multirow{3}{*}{1.3}&0.01347&1.0089\\
																	&                     &0.06685&1.0058\\
																	&										&0.08856&1.0512\\ 
																	\cline{3-4}
																	&\multirow{3}{*}{2.4}&0.01118&1.0089\\
																	&                    &0.06596&1.0337\\
																	&                    &0.09330&1.1509\\ 
		\hline
		\multirow{2}{*}{Trigonal(III)}&\multirow{2}{*}{3.1} & 0.06279 &1.0286 \\
		                               &                    &0.07903 & 1.0233\\

		\hline\hline
	\end{tabular}
\end{table}

\begin{figure}
	\includegraphics[scale = 0.7]{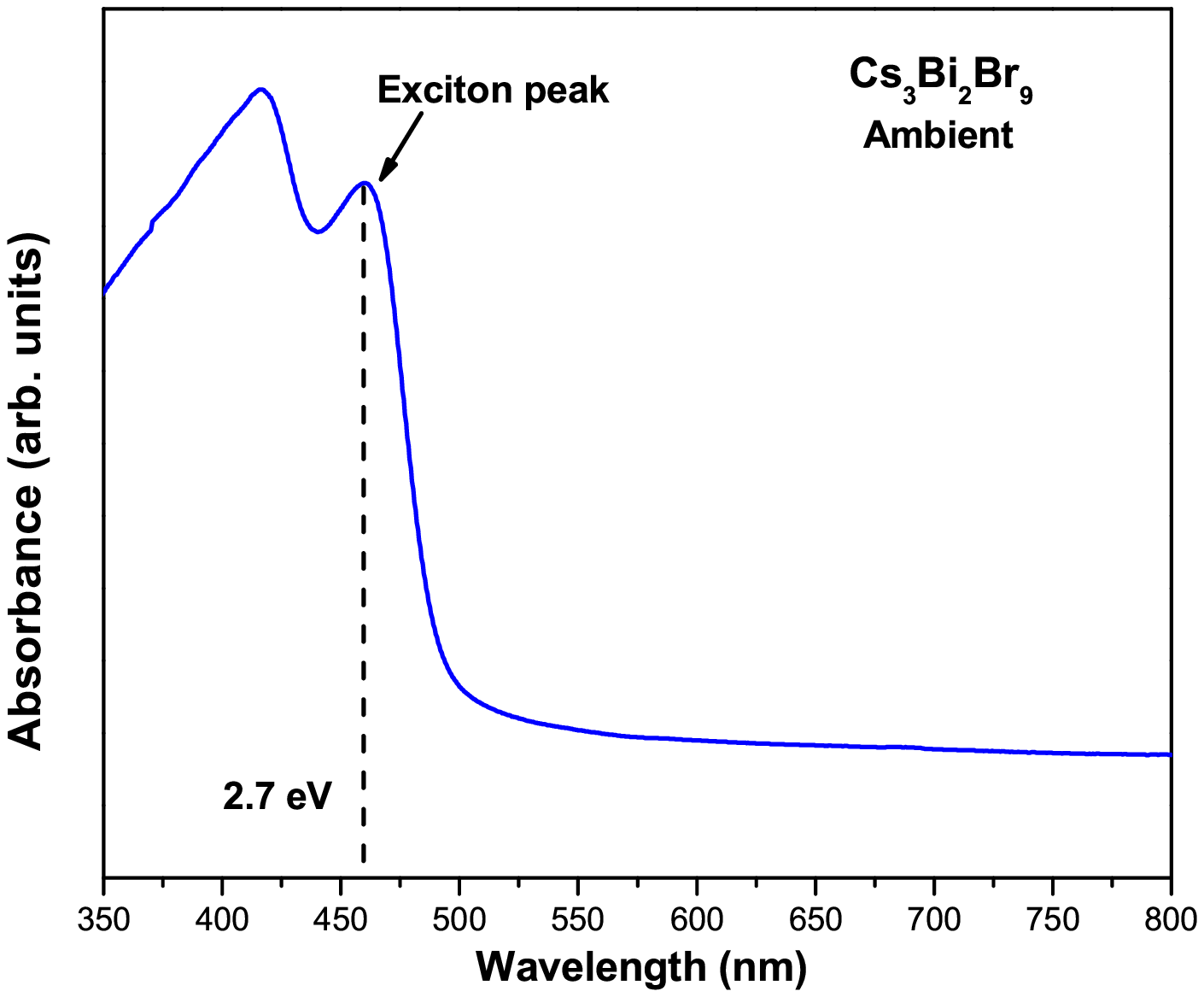}
	\caption*{Fig.S1: UV-absorption spectra of $Cs_3Bi_2Br_9$ at ambient condition. UV-absorption spectra shows an exciton peak at 2.7 eV  and an optical absorption edge at 2.5 eV.}
	
\end{figure}

\begin{figure}
	\centering
	\includegraphics[scale = 0.7]{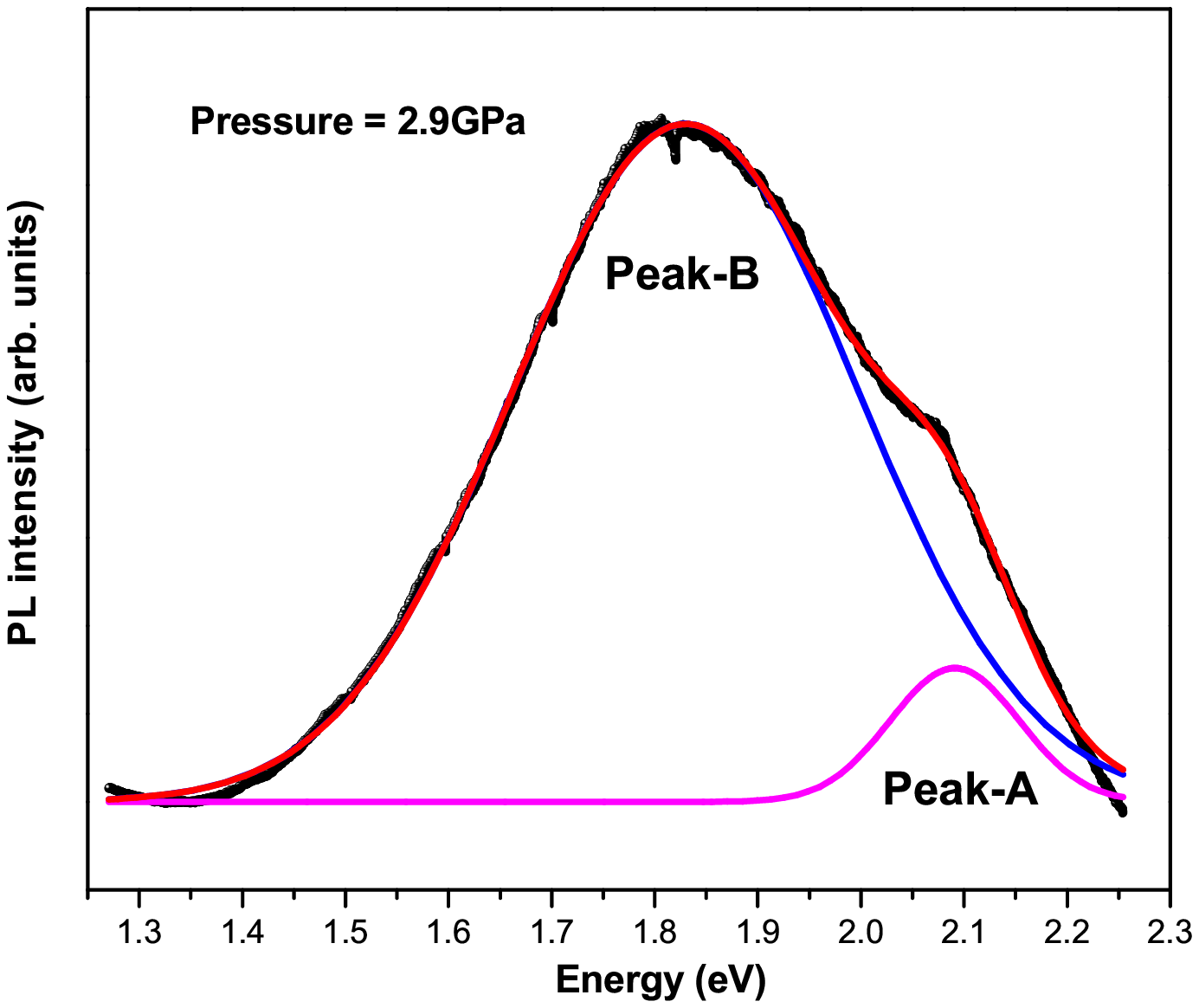}
	\caption*{Fig.S2: PL spectra of $Cs_3Bi_2Br_9$ at 2.9 GPa. Black dots represent observed data point. The blue, magenta  lines correspond to the fitting of  observed data point to two Gaussian  functions. Red line shows the sum of the two Gaussian fits. }
\end{figure}

\begin{figure}
	\centering
	\includegraphics[scale = 0.7]{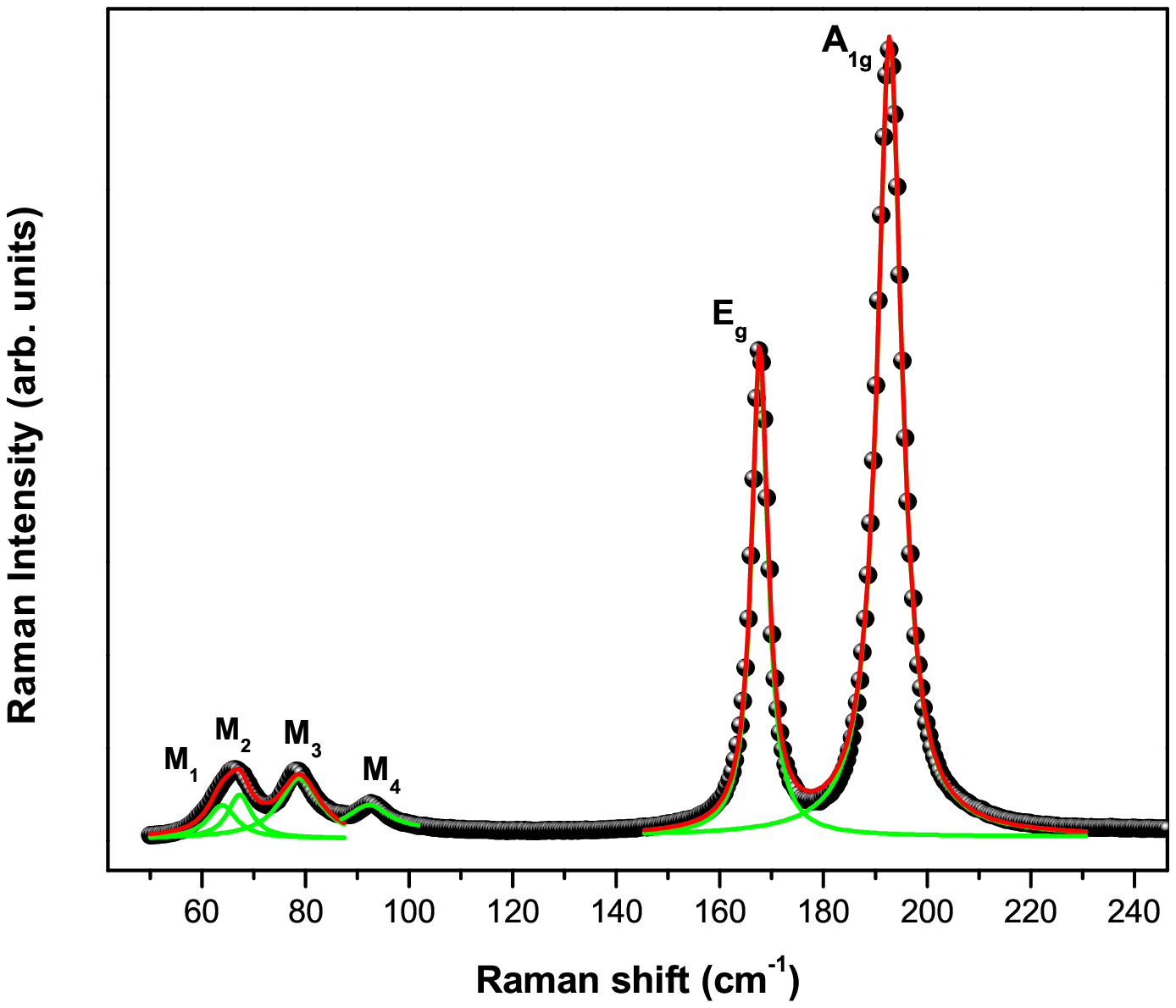}  		
		\caption*{Fig.S3: Deconvolution of ambient Raman spectra of $Cs_3Bi_2Br_9$ using Lorentzian function. Black dotes represent observed data point. The green line represent fits to individual peaks and red line represents total fit.}	
\end{figure}

\begin{figure}	
	\centering
	\includegraphics[scale = 0.5]{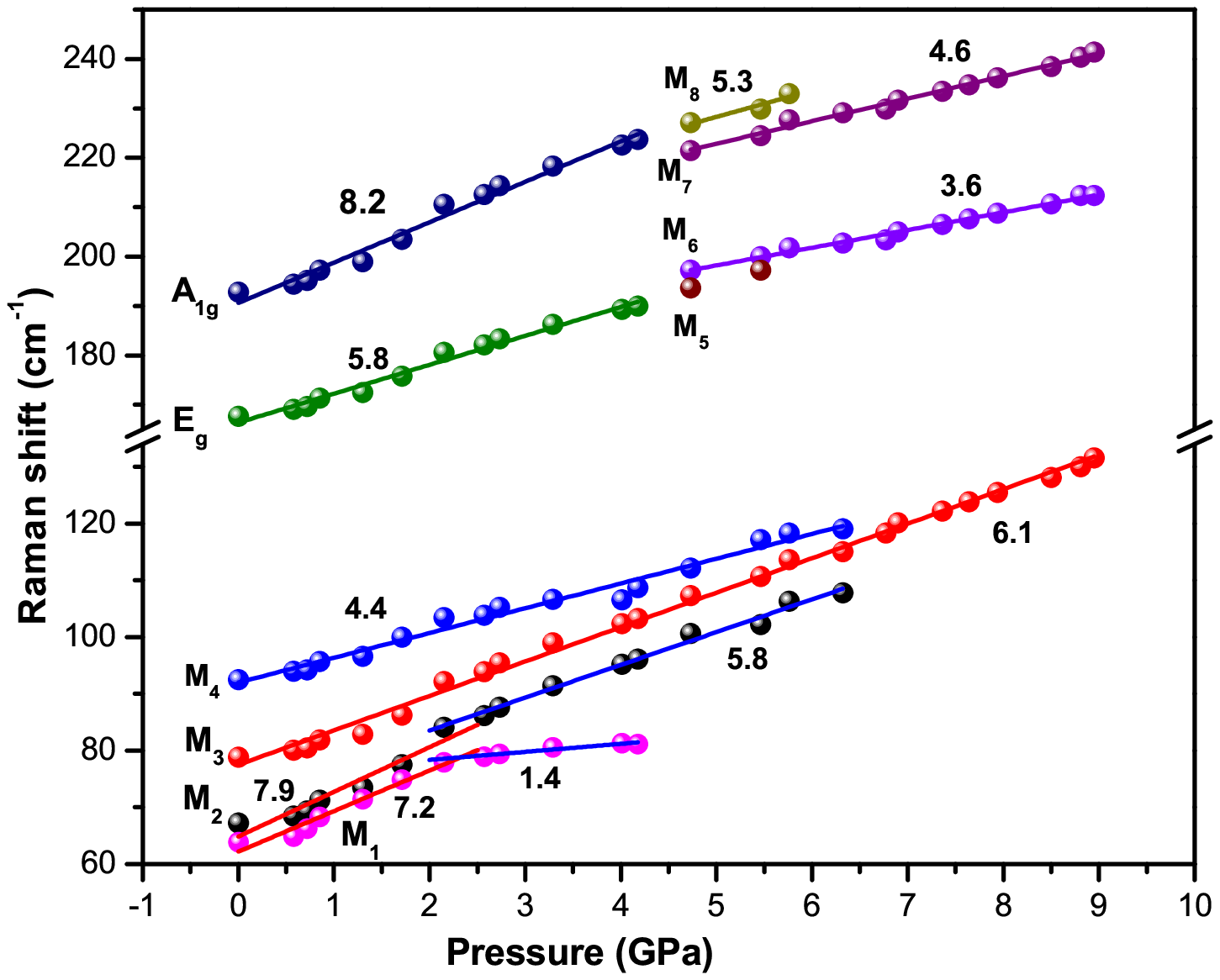}  	
		
	\caption*{Fig.S4: Variation of Raman mode frequencies with pressure. The slope of linear fitting of each mode with respect to pressure is shown. $M_1$ and $M_2$ modes also show a slope change at about 2.4 GPa.  }
\end{figure}

\begin{figure}
	\centering
	\includegraphics[scale=0.7]{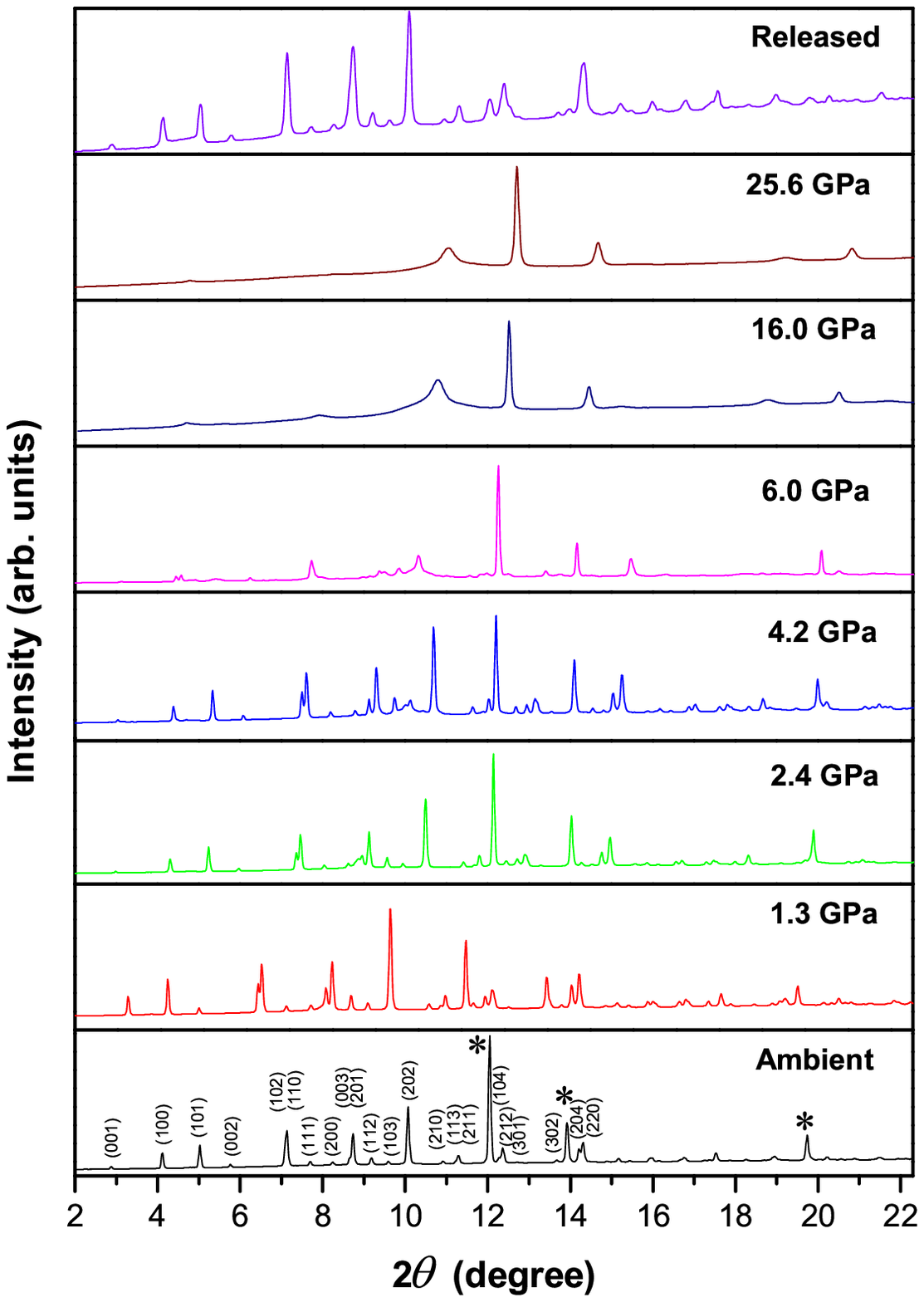}
	\caption*{Fig.S5: Pressure evolution of XRD patterns of $Cs_3Bi_2Br_9$ at selected pressure values. Stars indicate($\star$) Bragg peaks of pressure marker(Ag). }
\end{figure}

\begin{figure}
	\centering
		\begin{subfigure}[b]{0.49\textwidth}
			\includegraphics[scale=0.6]{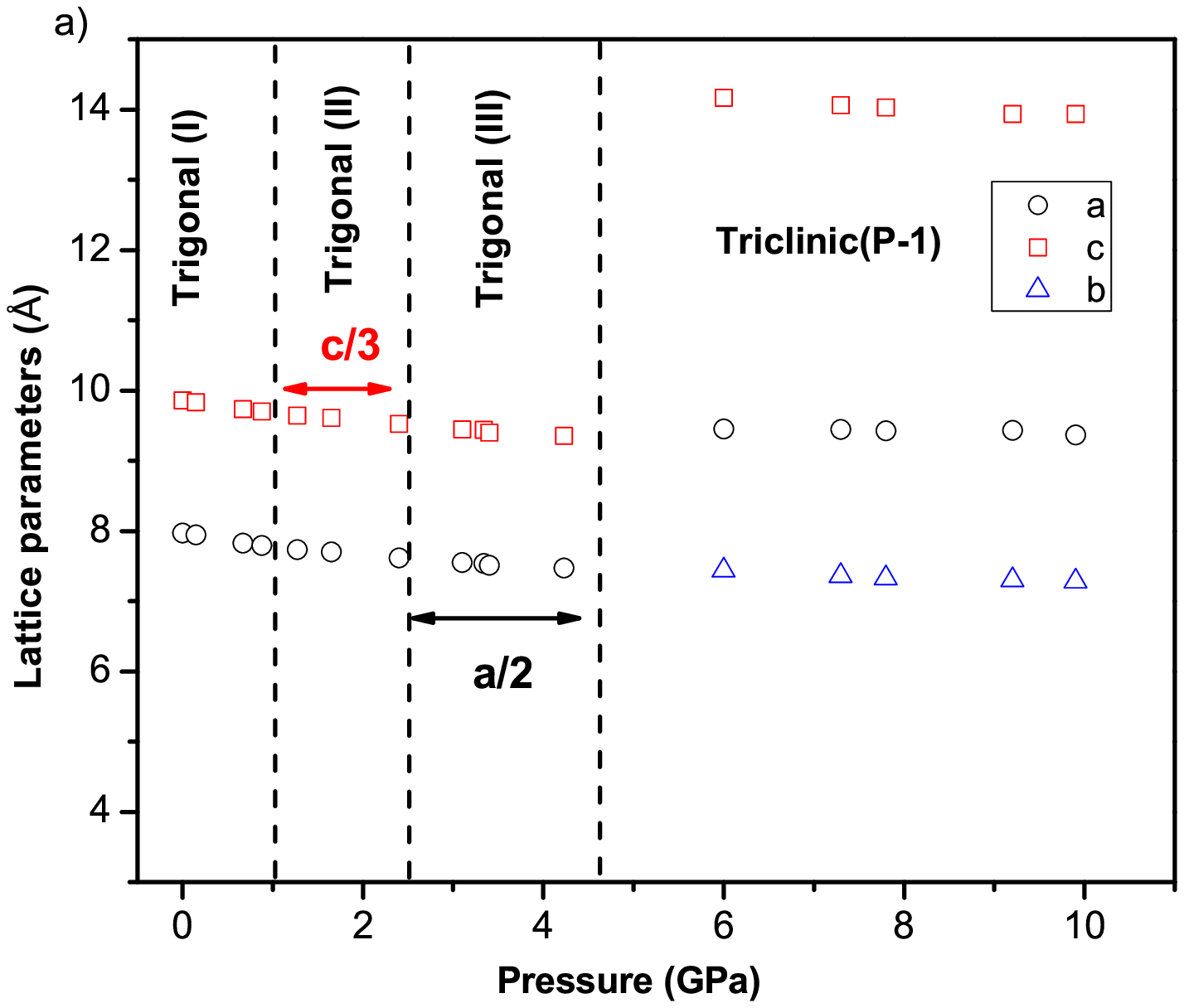}
		\end{subfigure}   
		
		\vspace*{-40mm}
		
		\begin{subfigure}[b]{0.49\textwidth}
			\includegraphics[scale=0.6]{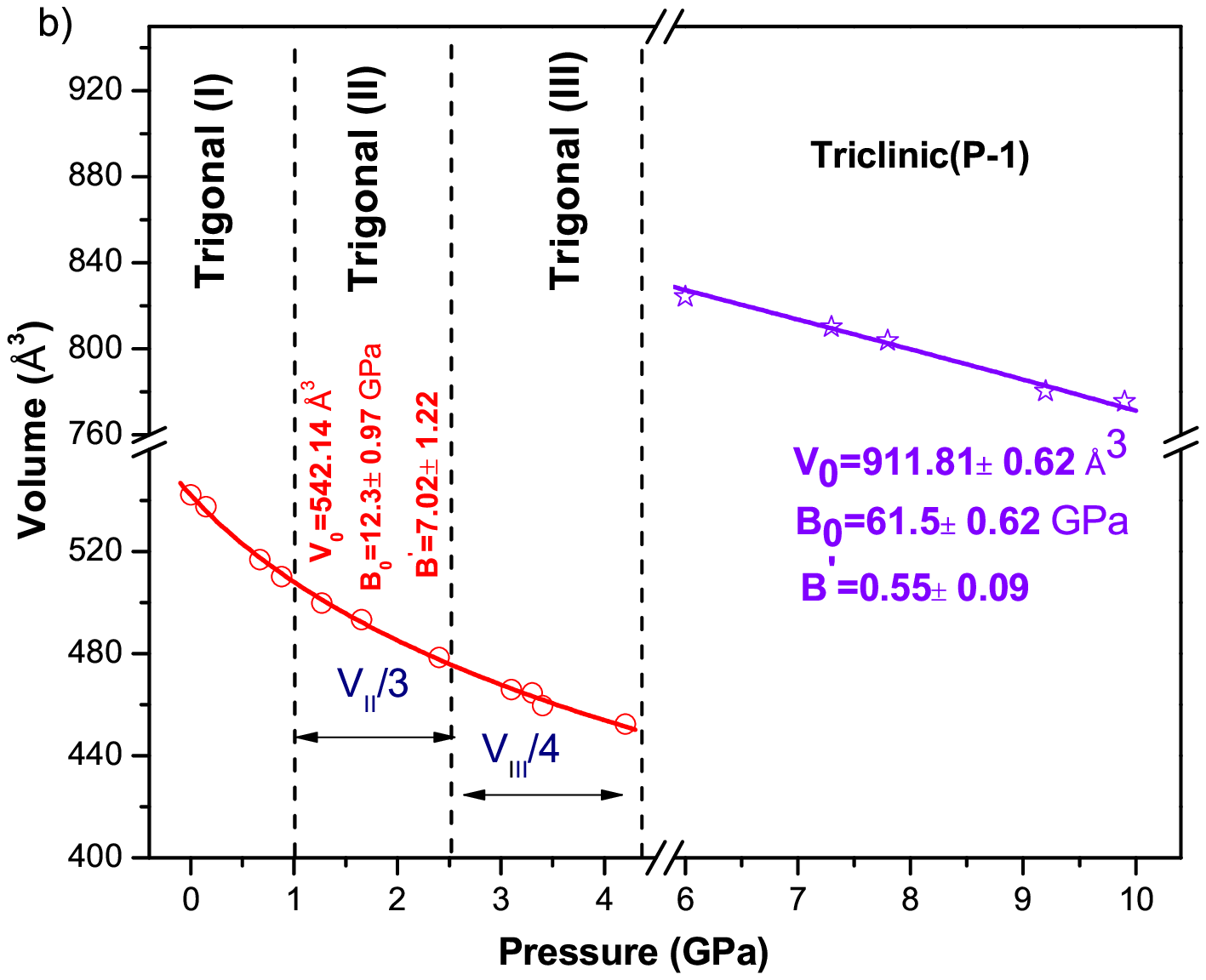}
		\end{subfigure}  		
		
	\vspace*{-20mm}	
		
		\caption*{Fig.S6: (a)Variation of normalized lattice parameters with pressure. (b) Normalized unit cell volume as a function of pressure. Observed data points are fitted to $3^{rd}$ Brich Murnaghan euqation of state. The solid lines show that fit. }
\end{figure}

\begin{figure}
\centering
  \begin{subfigure}[b]{0.49\textwidth}
    \includegraphics[scale=0.6]{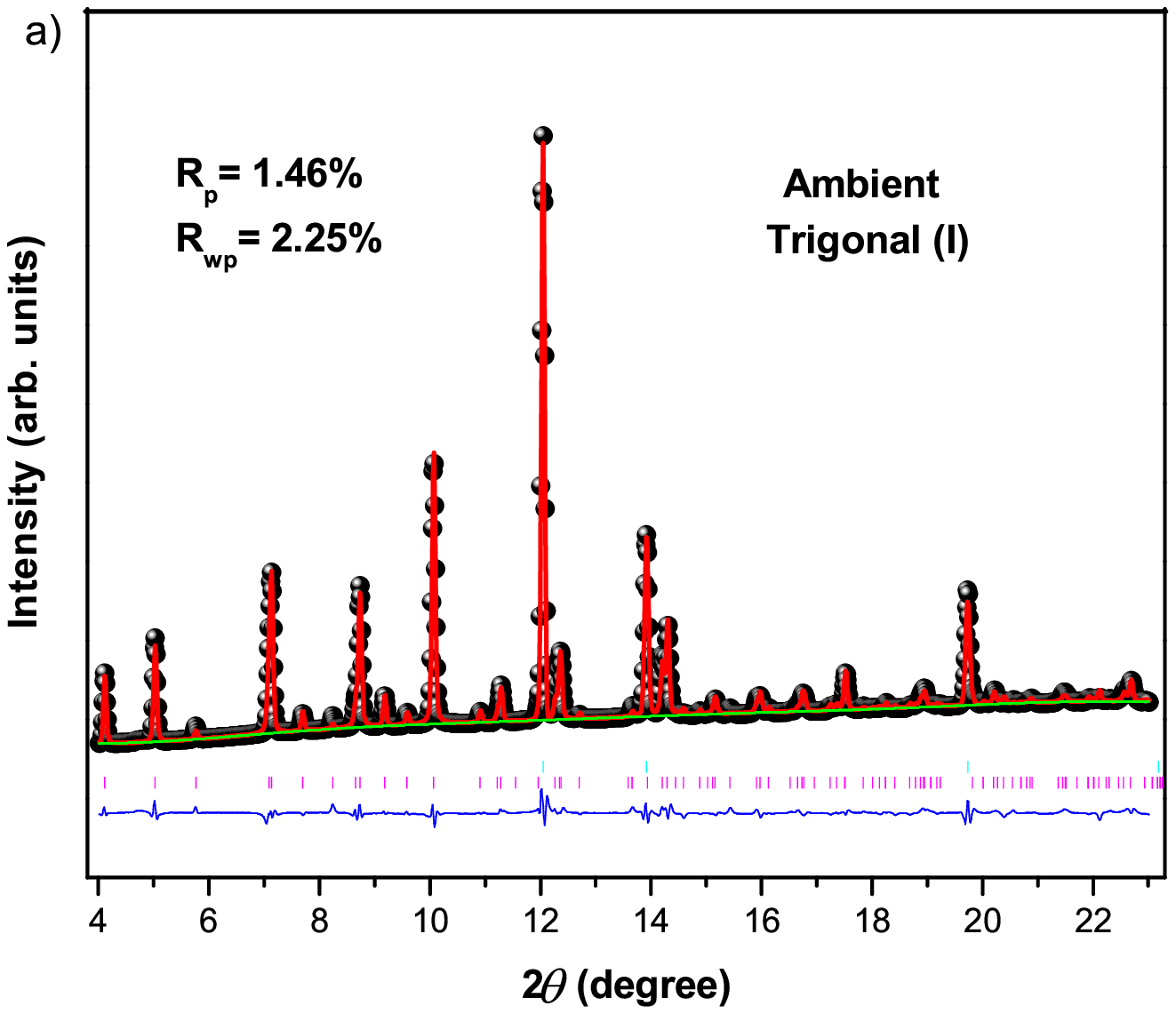}
  \end{subfigure}
  \begin{subfigure}[b]{0.49\textwidth}
    \includegraphics[scale=0.6]{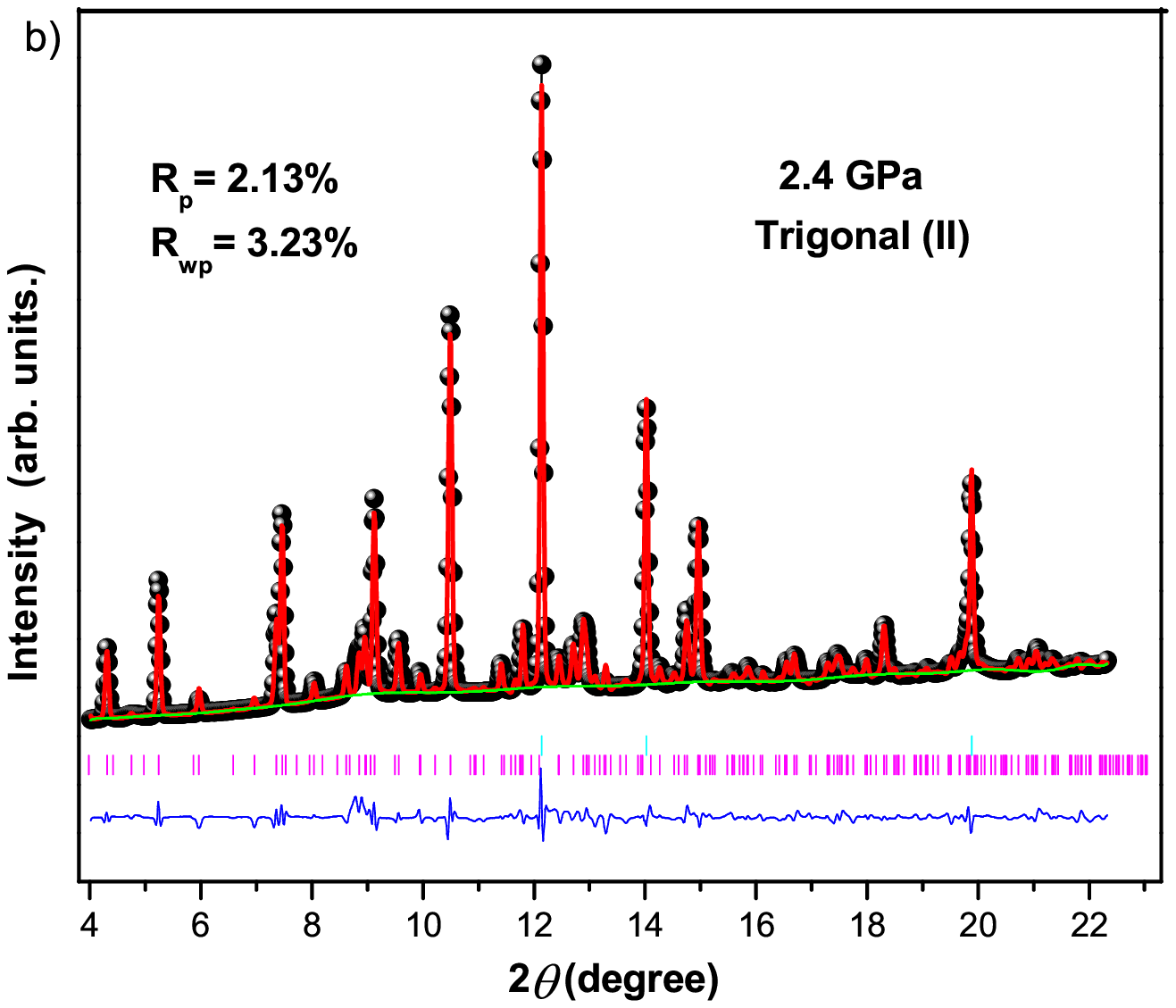}
  \end{subfigure}
	
	\vspace*{-40mm}
	
		\begin{subfigure}[b]{0.49\textwidth}
    \includegraphics[scale=0.6]{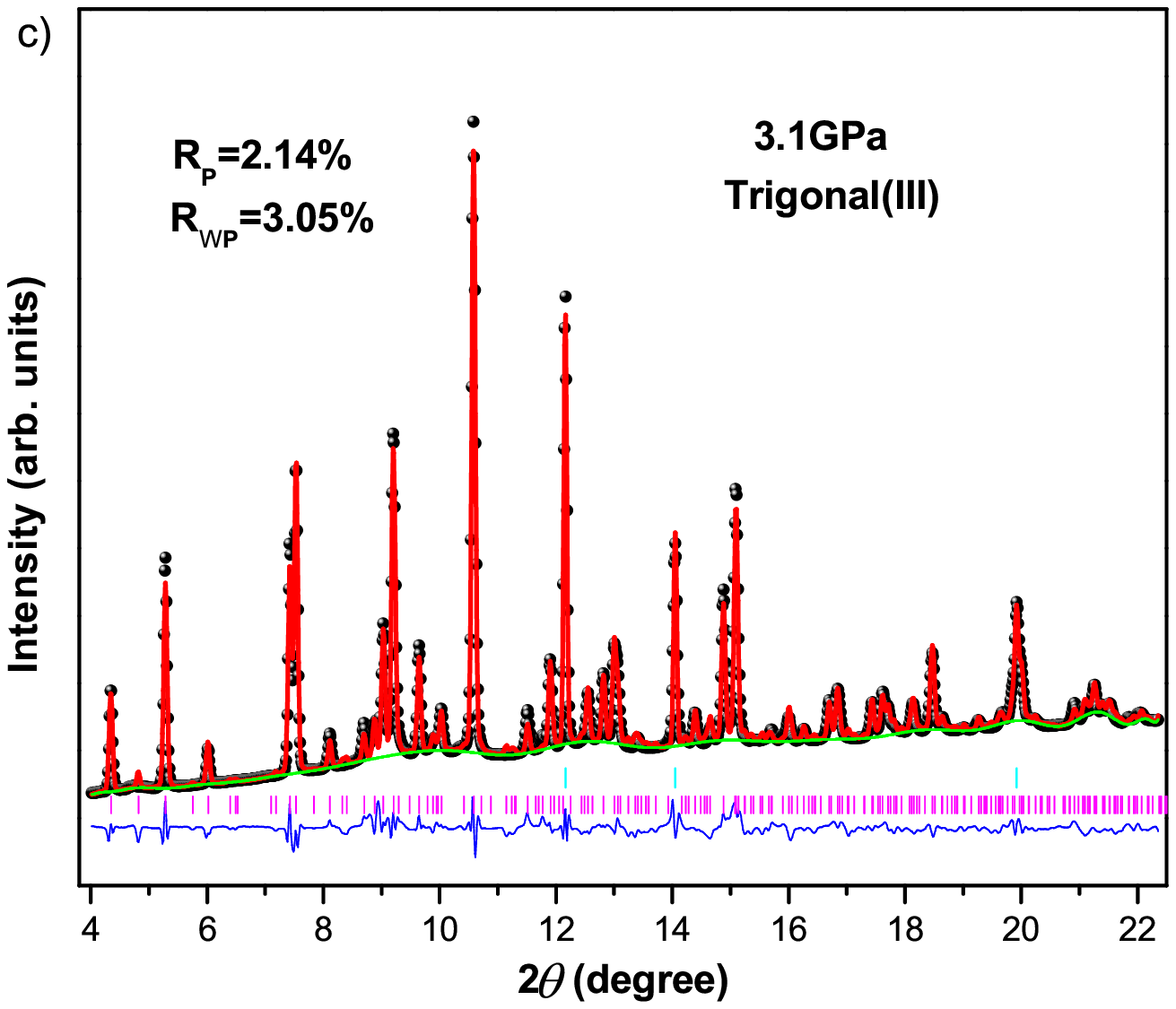}
  \end{subfigure}
  \begin{subfigure}[b]{0.49\textwidth}
    \includegraphics[scale=0.6]{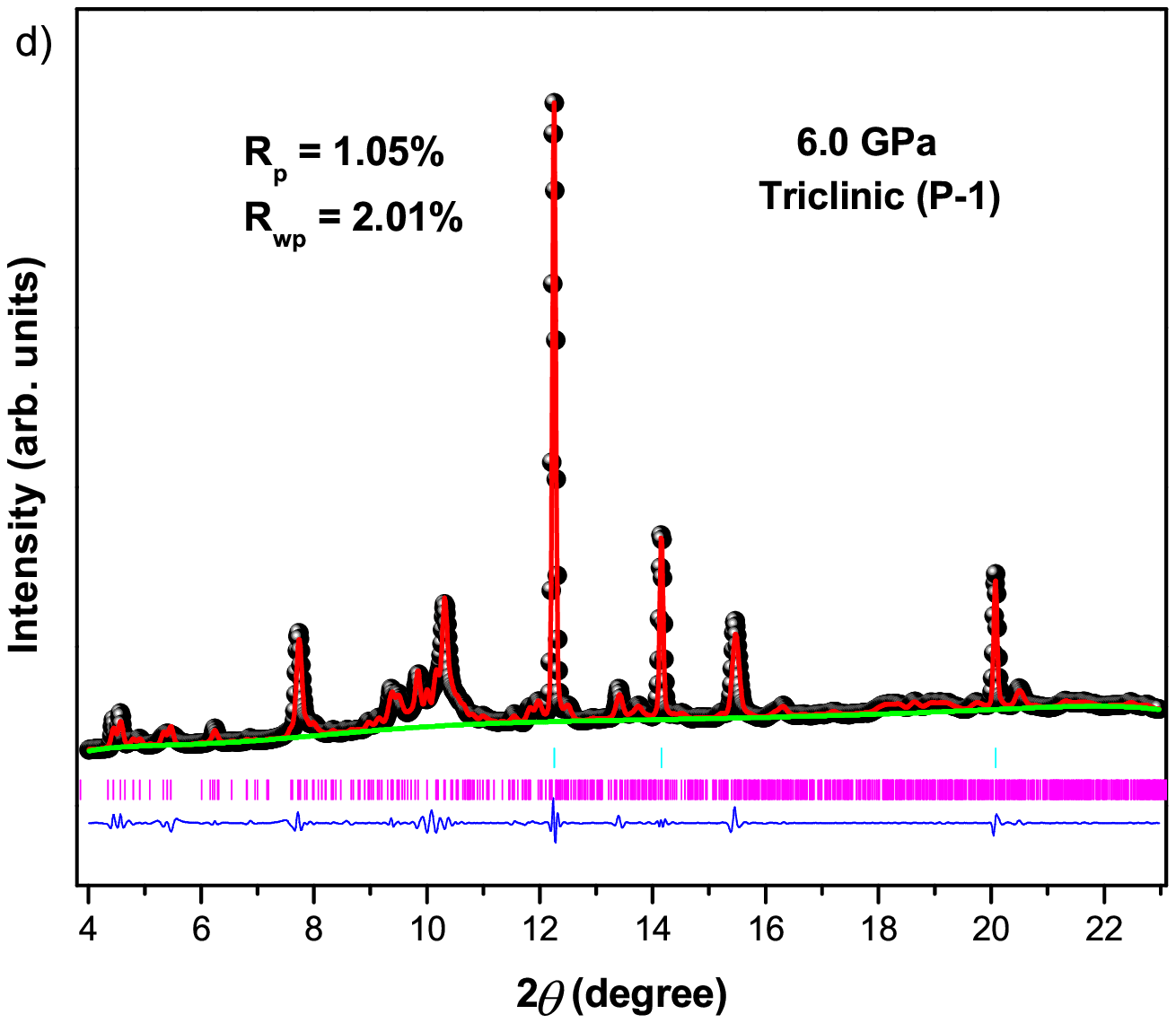}
  \end{subfigure}
	
\vspace{-40mm}

	\caption*{Fig.S7: Rietveld refinement obtained at (a) ambient condition, (b) 2.4 GPa, (c) 3.1 GPa and  (d) Lebail refinement at 6.0 GPa  of X-ray diffraction patterns of $Cs_3Bi_2Br_9$. black dotes show experimental data points. The red line fits to experimental data points. Background and difference are shown by green and blue line respectively. Magenta and Cyan vertical ticks indicate position of Bragg peaks of sample and pressure marker(Ag) respectively.}
	
\end{figure}

\begin{figure}
	\includegraphics[scale = 0.7]{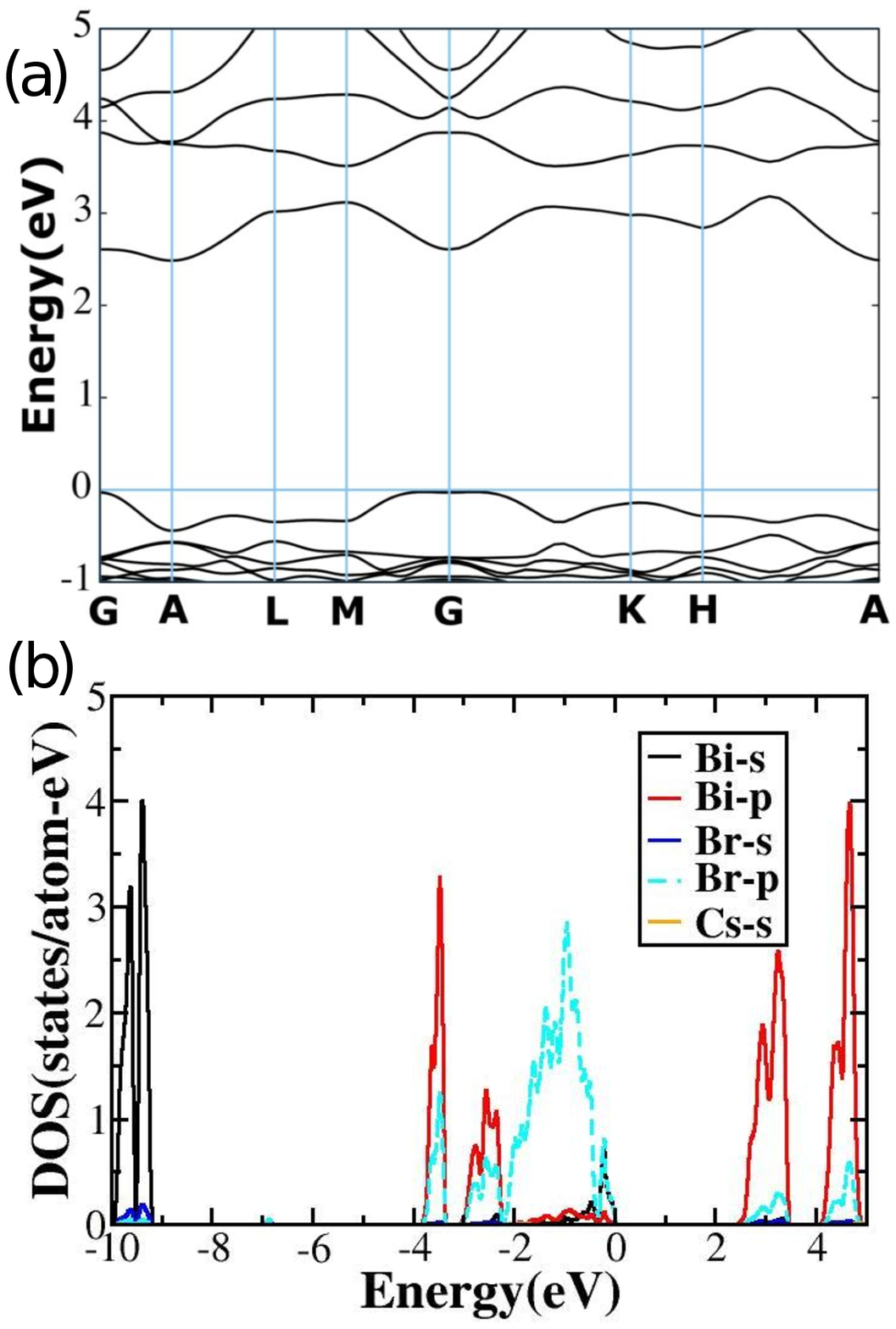}
	\caption*{Fig.S8: Calculated band structure and density of state  of $Cs_3Bi_2Br_9$ at 0.1GPa.}
	
\end{figure}

\begin{figure}
	\includegraphics[scale = 0.7]{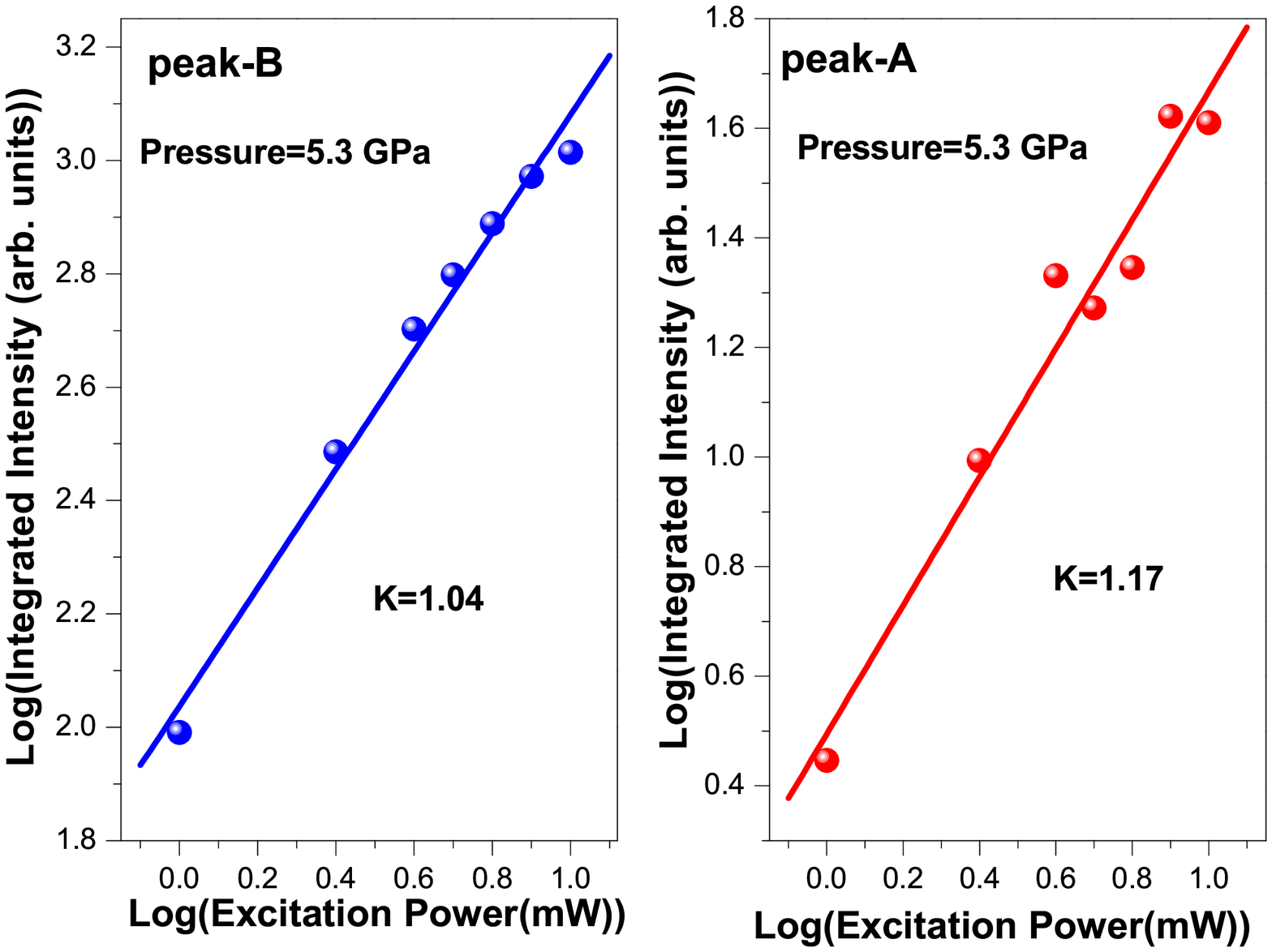}
	\caption*{Fig.S9: The obtained PL intensity ($I$) and excitation laser power ($L$)  follow the power law behaviour $I\sim L^K$ law. The obtained K values are shown in figure.}
	
\end{figure}